\documentstyle[aps,prl,floats,epsf]{revtex}
\begin{document}
\renewcommand{\theparagraph}{\Alph{paragraph}}
\draft
\twocolumn[\hsize\textwidth\columnwidth\hsize\csname@twocolumnfalse%
\endcsname

\title{A Model for Nonequilibrium Wetting Transitions in Two
Dimensions\\}

\author{H. Hinrichsen$^{(1,2)}$, R. Livi$^{(3)}$, 
        D. Mukamel$^{(1)}$, and A. Politi$^{(4)}$\\[-2mm]$ $}

\address{$^{1}$
        Department of Physics of Complex Systems,
        Weizmann Institute, Rehovot 76100, Israel\\
        $^{2}$
        Max-Planck-Institut f\"ur Physik komplexer Systeme,
        N\"othnitzer Stra\ss e 38, 01187 Dresden, Germany\\
        $^{3}$
        Dipartimento di Fisica, Universit\'a, INFM and INFN, 50125 
        Firenze, Italy\\
        $^{4}$
        Istituto Nazionale di Ottica and INFN-Firenze, 50125 Firenze,
        Italy}

\date{August 18, 1997, submitted to {\it Phys. Rev. Lett.}}

\maketitle
\begin{abstract}
A simple two dimensional ($2d$) model of a phase growing on a substrate
is introduced. The model is characterized by an adsorption rate $q$,
and a desorption rate $p$. It exhibits a wetting transition which may be
viewed as an unbinding transition of an interface from a wall.
For $p=1$, the model may be mapped onto an exactly soluble 
equilibrium model exhibiting complete wetting with critical exponents
$\gamma = 1/3$ for the diverging interface width and $x_0 = 1$ for
the zero-level occupation. For $0 < p \neq 1$ a crossover to different
exponents is observed which is related to a KPZ type nonlinearity.
\end{abstract} \vspace{2mm}
]
%
%
%
%
%
%
The interaction of a bulk phase $(\alpha )$ of a system with a wall
or a substrate may result in very interesting wetting phenomena. In
particular a layer of a second phase~$(\beta )$, which is preferentially
attracted to the wall may be formed in its vicinity. As some of the
parameters controlling the system, say temperature or chemical
potential, are varied, the 
thickness of the $\beta$ layer may diverge, leading to a 
wetting transition. Such transitions have been theoretically studied
and experimentally observed in a variety of systems and models
in thermal equilibrium (for a review, see Ref.~\cite{WettingReview}).
Wetting transitions may be viewed as the unbinding of an
interface from a wall. Within this approach one considers
the interface configuration $h(r)$ which gives the height of the
interface above the wall at point $r$. 
One then introduces an effective Hamiltonian
of the form \cite{EquilibriumFieldTheory}
\begin{equation}
\label{Hamiltonian}
{\cal H} = \int d^{d-1}r [ \frac{\sigma}{2} (\nabla h)^2 + V(h(r)) ] \,,
\end{equation}
where $\sigma$ is the surface tension of the $\alpha$-$\beta$ interface,
$V(h(r))$ yields the effective interaction between the wall and the
interface, and $d-1$ is the interface dimension.
The potential $V$ which contains an attractive component may
bind the interface to the wall. 
However, as the temperature or other parameters describing
the system are varied, the attractive component of the potential
may become weaker and it is no longer able to bind the interface,
leading to a wetting transition.
In $d=2$ dimensions one usually distinguishes between critical wetting
and complete wetting. Critical wetting is marked by the divergence of
the interface width when the temperature $T$ is 
increased towards the transition
temperature $T_W$ moving along the coexistence curve of the $\alpha$ and
$\beta$ phases. On the other hand complete wetting is characterized by the
divergence of the interface width when the chemical potential difference
between the two phases is varied, moving towards the coexistence curve
at $T > T_W$. These types of transitions are associated with two
different sets of critical exponents.

A very interesting question which has not been studied in detail so 
far is that of wetting transitions under {\em non-equilibrium}
conditions. Here the $\beta$ phase is adsorbed to the wall via
a growth process whose dynamics, unlike that of equilibrium
processes, does not obey detailed balance.
This problem may be studied by considering the 
behavior of a moving interface interacting with a wall.
Such transitions have been reported in recent studies of the dynamics
of certain models of coupled maps \cite{CoupledMaps}.

In this Letter we introduce a class of nonequilibrium growth models of 
a one dimensional interface interacting with a substrate. The 
interface evolves by both adsorption and desorption processes which in
general do not satisfy detailed balance. By varying the
relative rates of these processes, a transition from a binding to a
non-binding phase is found. For a particular value of the desorption
rate, for which the dynamics happens to have detailed balance, 
the model may be mapped onto an exactly soluble equilibrium model
which exhibits a complete wetting unbinding transition. The 
associated critical exponents are $\gamma = 1/3$ for
the interface width and $x_0 = 1$ for the base level occupation.
For generic values of the desorption rate, however, detailed balance 
is violated and a crossover to different exponents is observed.

\paragraph{Definition of the model: }
The model is defined in terms of growth of a $1d$ interface,
in which both adsorption and desorption processes take place. We
consider a {\em restricted} solid-on-solid (RSOS) growth process, 
where the height differences between neighboring sites are
restricted to take values $0,\pm1$. 
The model is defined on a 1$d$ lattice of $N$ sites
with associated height variables $h_i=0,1,\ldots,\infty$ and periodic
boundary conditions. We use random sequential dynamics which are
defined through the following algorithm: at each update choose a site
$i$ at random and attempt to carry out one of the processes

-  adsorption of an adatom with probability $q/(q+p+1)$:
        \begin{equation} \label{Process1}
        h_i \rightarrow h_i+1 \end{equation}
        
- desorption of an adatom from the edge of an island
        with probability $1/(q+p+1)$:
        \begin{equation} \label{Process2} 
        h_i \rightarrow \min(h_{i-1},h_{i},h_{i+1})$$
        \end{equation}
        
- desorption of an adatom from the interior of an island
        with probability $p/(q+p+1)$:
        \begin{equation}  \label{Process3}
        h_i \rightarrow h_i-1 \quad \mbox{if} \quad
        h_{i-1}=h_{i}=h_{i+1} \end{equation}
If the selected process would result in a violation of the RSOS
constraint $|h_i-h_{i+1}| \leq 1$,
the attempted move is abandoned and a new site $i$ is selected.
In addition, a hard-core wall at zero height is introduced, 
i.e. a process is only carried  out 
if the resulting interface heights are non-negative.
One can prove that these processes 
in general do not satisfy detailed balance.

The presence of a hard-core wall at $h=0$ 
leads to a phase transition that takes
place even in finite systems. This can 
be seen as follows. Without the wall
the interface in a finite system has 
a finite width. For fixed $p>0$ the
parameter $q$ controls the mean growth 
velocity of the interface, i.e. for
large $q$ the interface grows while for small $q$ it moves
downwards. These two regimes are separated by a critical 
growth rate $q=q_c$ for which
the mean velocity is zero. Therefore, 
on large time scales, a lower wall will 
only affect the interface dynamics 
if the interface does not move away from
the wall, i.e. $q \le q_c$, resulting
in a smooth interface. In the growing 
phase $q>q_c$, however, the interface
does not feel the wall. It is rough and propagates 
with a constant mean velocity. The phase
transition line for an infinite system 
is shown in Fig.~1.

\begin{figure}
\epsfxsize=70mm 
\begin{center}~ 
\epsffile{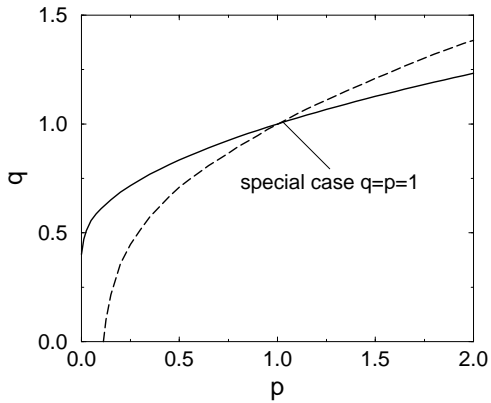}
\end{center}
\caption{
Phase diagram for an infinite system.  
The wetting transition takes place along the solid line.
Along the dashed line 
the growth velocity in the model without a wall does not depend
on a global tilt of the interface, indicating that the effective
KPZ nonlinearity vanishes. Both lines intersect in the point $q=p=1$, 
where the microscopic processes obey detailed balance.}
\end{figure}

Throughout this paper we are particularly interested in the
the mean growth velocity $v$ in the growing phase, 
the occupation $\rho_0$ of the zero-height layer in the smooth phase,
and the interface width in the smooth phase which is defined by
\begin{equation}
w^2 = \frac{1}{N} \sum_{i=1}^N \Bigl(h_i-\frac{1}{N} 
\sum_{j=1}^N h_j\Bigr)^2\,.
\end{equation}
Consider now the thermodynamic limit $N \to \infty$.
Near criticality in the growing phase $q>q_c$, we expect 
the interface velocity to scale like
\begin{equation}
v\sim(q-q_c)^y\,,
\end{equation}
whereas in the smooth phase, $q<q_c$, the expected scaling 
for bottom layer occupation and width is
\begin{equation}
\rho_0\sim(q_c-q)^{x_0}\,, \quad\quad
w\sim(q_c-q)^{-\gamma}\,.
\end{equation}
It would be interesting to find out how the critical exponents
$y$, $x_0$, and $\gamma$ depend on $p$. We note that the case $p=0$
is special: In this case atoms cannot be desorbed from a completed
layer, and the interface cannot move below its actual minimum height. 
This means that the hard-core wall becomes irrelevant.
Therefore, the phase transition (which still exists for $p=0$)
relies on a completely different mechanism.
The $p=0$ transition is expected to belong to the universality class
of a closely related model previously considered in Ref.~\cite{OurPaper}.
It has been shown that in this case some of the 
critical exponents can be related to
the universality class of directed percolation (DP). In particular
one expects $x_0 = \beta$ and $y =  \nu_{\perp}$, where $\beta = 0.276$
and $\nu_{\perp}=1.73$ are the density and correlation length exponents,
respectively, of DP. Here
completed layers play the role of absorbing states from
where the system cannot escape. 
For $p>0$, however, the system is ergodic 
and one expects different critical exponents.
Another special case is $p=1$: here the system does
satisfy detailed balance and can be solved exactly. In the following, we
present our analysis of the wetting transition for $p=1$. 
We then consider the general case for which 
the model does not have detailed balance.

\paragraph{Exactly soluble case, $p=1$:}
We first show that in this case
the steady state satisfies detailed balance and that for $q <1$ the 
probability of finding the interface in a particular configuration
$\sigma_H=\{h_1,\ldots,h_N\}$ is given by the distribution
\begin{equation}
\label{EquilibriumDist}
P(h_1,\ldots,h_N) = P(\sigma_H) = Z_N^{-1} \,  q^{H(h_1,\ldots,h_N)}\,,
\end{equation}
where 
\begin{equation}
H=H(h_1,\ldots,h_N)=\sum_{i=1}^Nh_i
\end{equation}
is the sum of all heights.
Here, the partition sum $Z_N=\sum_{h_1,\ldots,h_N} q^{H}$ 
runs over all interface configurations which respect the RSOS
and the hard-core wall constraints.

In order to prove Eq.~(\ref{EquilibriumDist}), notice that
the processes (\ref{Process1})--(\ref{Process3}) subjected to
the RSOS constraint correspond to a change of $H$ by one unit. 
Therefore, if the distribution of Eq.~(\ref{EquilibriumDist}) is
to hold in the steady state, the probabilities 
of finding the interface in states with total
height $H$ and $H+1$ have to satisfy
\begin{equation}
\label{ProbabilityQuotient}
P(\sigma_{H+1})/P(\sigma_H) = q\,.
\end{equation}
As one can read off from the processes
(\ref{Process1})--(\ref{Process3}), 
any allowed transition $\sigma_H\rightarrow\sigma_{H+1}$ occurs with
rate $w(\sigma_H \rightarrow \sigma_{H+1})=q$. 
Moreover, for each such process $\sigma_H \rightarrow \sigma_{H+1}$, 
there is a reverse process $\sigma_{H+1}\rightarrow \sigma_H$ 
which takes place with a nonzero rate. 
For $p=1$, this rate is given by
$w(\sigma_{H+1}\rightarrow \sigma_H)=1$ so that
\begin{equation}
w(\sigma_H \rightarrow \sigma_{H+1}) /
w(\sigma_{H+1} \rightarrow \sigma_H) = q\,. \quad \quad (p=1)
\end{equation}
Together with Eq.~(\ref{ProbabilityQuotient}) 
this implies that the processes
(\ref{Process1})--(\ref{Process3})
satisfy detailed balance. Notice that the above
consideration is consistent with the
hard wall constraint $h_i \geq 0$. Equation~(\ref{EquilibriumDist}) yields
the steady state distribution only for $q < 1$. The unbinding transition
takes place at $q=1$ and the height distribution becomes time dependent
for $q > 1$.
For $p \neq 1$, detailed balance is violated. This can be proven
by construction of explicit cycles of configurations 
in small systems for which the rates
of moving clockwise and counter-clockwise are unequal.

We now apply the transfer matrix formalism \cite{Hilhorst,Burkhardt}
to study the distribution~(\ref{EquilibriumDist}).
Let us define a transfer matrix~$T$ acting in spatial direction by
\begin{equation}
\label{TransferMatrixDefinition}
T_{h,h'} = \left\{
\begin{array}{ccl}
q^{h} && \mbox{if} \, |h-h'| \leq 1 \\
0 && \mbox{otherwise} 
\end{array} \right.\,,
\end{equation}
where $h,h'\geq 0$. Steady state properties can be derived from the
eigenvector $\phi$ that corresponds to the largest eigenvalue $\mu$
of the transfer matrix
$\sum_{h'=0}^\infty T_{h,h'}\phi_{h'} = \mu\,\phi_h$. From 
the squares of the eigenvector components one can derive various
steady state quantities. For example, 
the probability $\rho_h$ of finding the interface at height $h$
is given by $\rho_h=\phi_h^2/\sum_{h'}\phi_{h'}^2$. 
Here we are particularly interested in the scaling behavior
of bottom layer occupation $\rho_0$ and the width
$w^2=\sum_h(h-\bar{h})^2\rho_h$, where  
$\bar{h}=\sum_hh\rho_h$ denotes the mean height.

Close to criticality, where $\epsilon=1-q$ is small, one can carry out
the continuum limit $\phi_h \rightarrow \phi(\tilde{h})$, replacing the
discrete heights $h$ by real-valued heights $\tilde{h}$. Then,
the above eigenvalue problem turns into a differential equation
\cite{Hilhorst} which, to leading order in $\epsilon$, is given by
\begin{equation}
\label{DifferentialEquation}
\Bigl( \frac{\partial^2}{\partial \tilde{h}^2} + (3-\mu) - 
3 \epsilon \tilde{h} \Bigr)\, \phi(\tilde{h}) = 0\,.
\end{equation}
%
%
\begin{figure}
\label{FigureMC}
\epsfxsize=90mm 
\epsffile{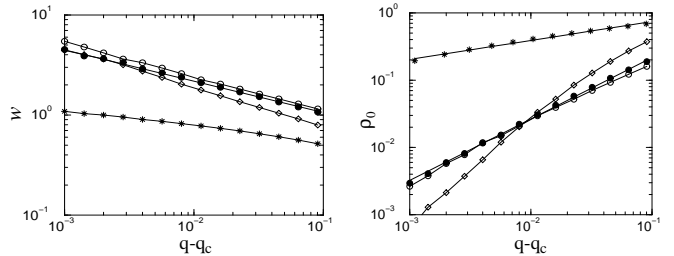}
\caption{
Results obtained from Monte Carlo simulation of the growth model 
with 1500 sites. The width $w$ and the bottom layer occupation
$\rho_0$ are measured for ($\star$)~$p$=0, ($\diamond$)~$p$=0.05,
($\bullet$)~$p$=1, and ($\circ$)~$p$=2.0. For $p$=1 we fitted straight
lines. 
The curvature for $p\neq 1$ indicates a crossover to KPZ exponents.
}
\end{figure}

This equation, together with the boundary conditions
$\phi(0)=\phi(\infty)=0$, has a unique physical solution. Simple dimensional
analysis indicates that the height
variables scale as $h \sim \epsilon^{-1/3}$ and thus the
width diverges as $w^2 \sim \epsilon^{-2/3}$.  The occupation
of the bottom layer in the continuum limit is given by
$\rho_0 = {\cal N}^{-1} (\phi'(0))^2$, where ${\cal N} =
\int d\tilde{h} \phi^2(\tilde{h})$ is a normalization factor.
Since $\phi'(0) \sim \epsilon^{1/3}$ and ${\cal N} \sim \epsilon^{-1/3}$
one obtains a linear scaling law $\rho_0 \sim \epsilon$. The critical
exponents for $p=1$ are thus given by
\begin{equation}
\label{EWCriticalExponents}
x_0=1\,,\quad\quad
\gamma=1/3\,.
\end{equation}

\paragraph{Numerical results:}
In order to determine the critical exponents for other values
of the growth rate $p$, 
we perform Monte-Carlo simulations.
The width $w$ and the occupation
of the bottom layer $\rho_0$ are measured in the smooth phase $q < q_c$.
Depending on $\epsilon=q_c-q$, we 
first equilibrate a system of size $1500$
over a time interval up to $4 \times 10^6$ 
time steps. Then the thermal 
averages of $w$ and $\rho_0$ are measured over a time interval 
of the same size. Similarly, the interface 
velocity $v$ is measured in the 
growing phase $q>q_c$.

The numerical data measured in the smooth phase
are shown in Fig.~2. From the 
slopes in the double logarithmic plots 
we estimate the critical exponents (see Table~I).
For $p=1$, the numerical results we obtain are consistent with
the exact values derived above. In addition the velocity exponent is
found to be $y=1.01(3)$.
For $0 < p < 1$ we observe a {\it crossover} to different
critical exponents. Since the values differ from those in 
Eq.~(\ref{EWCriticalExponents}) by less than 30\% and the crossover 
is extremely slow, it is difficult to determine these exponents 
precisely. Our estimates are $y=1.00(3)$, $x_0=1.5(1)$, and
$\gamma=0.41(3)$. Similar results are obtained for $p>1$ 
except for $x_0$ which is close to one in this case.
Finally, for $p \rightarrow 0$, we observe another crossover. 
This is consistent with the results of
Ref.~\cite{OurPaper} which indicate that the interface width
diverges logarithmically at the $p=0$ transition.
\begin{table}
\begin{tabular}{|c||c|c|c|c|}
           & $p=0$         & $p=0.05$       & $p=1.0$        & $p=2.0$
\\\hline
$q_c$      & $0.3991(5)$   & $0.5564(2)$    & $0.9999(2)$    &
$1.2329(5)$\\
$x_0$      & $0.27(1)$     & $1.51(6)$      & $0.96(5)$      & $1.02(5)$
\\
$\gamma$   & $0$ (log.)    & $0.41(3)$      & $0.32(3)$      & $0.37(3)$
\\
$y$        & $1.69(5)$     & $0.98(3)$      & $1.01(3)$      & $1.00(3)$
\\
\end{tabular}
\vspace{2mm}
\caption{Estimates for the critical exponents.}
\end{table}
\paragraph{General Considerations:}
Equilibrium wetting of a $2d$ Ising system
has been studied in Refs.~\cite{Abraham,Hilhorst}. It was shown that a
column of weak bonds (which acts as an attractive potential for the
domain wall separating the up and down states)
located at the boundary induces a wetting 
transition at some finite temperature $T_W$. 
At the critical wetting transition
the interface width diverges with $\gamma = 1$.

The transition found for $p=1$ is of different nature. In this
case, $(1-q)$ acts as a chemical potential difference between the two
coexisting phases. For $q<1$ the chemical potential difference drives
the interface towards the wall, resulting in a smooth interface.
On the other hand, for $q>1$ the interface is driven away from the wall,
resulting in a KPZ-like rough moving interface \cite{KPZ}. 
The critical behavior associated
with the transition is thus that of {\em complete} wetting.

For $p \neq 1$, the mapping to equilibrium is impossible since detailed
balance is violated. Here a KPZ-type nonlinearity is expected to be
responsible for the different exponents we observe. Within this
approach,
one describes the system by the Langevin equation
\begin{eqnarray}
\label{LangevinEquation}
\frac{\partial h(r,t)}{\partial t} &=& v_0 + \sigma\nabla^2 h(r,t) - 
\frac{\partial V(h(r,t))}{\partial h(r,t)} \\
&& \noindent + \lambda (\nabla h(r,t))^2 + \zeta(r,t)\,, \nonumber
\end{eqnarray}
where $\zeta(r,t)$ is a zero-average Gaussian noise field with variance
$\langle\zeta(r,t)\zeta(r',t')\rangle=2D\delta^{d-1}(r-r')\delta(t-t')$
and $V$ is the effective interaction between the wall and the interface.
This equation has been studied recently \cite{Geoff,Hwa} in the context 
of nonlinear diffusion with multiplicative noise. A simple scaling
argument
suggests that the width exponent corresponding to the wetting transition
described by this equation is given by 
\begin{equation}
\label{KPZExponent}
\gamma = (2-z)/(2z-2)\,,
\end{equation}
where $z$ is the dynamic exponent. 
For a 1d interface $z=3/2$, yielding $\gamma =1/2$. Our numerical results
indicate that the width exponent $\gamma$ is larger for $p \ne 1$
as compared with its $1/3$ value at $p=1$, although it seems to be smaller
than $1/2$. However, in view of the very slow crossover expected in this
problem (see below) it is possible that $\gamma$ is indeed $1/2$, but
more extensive simulations close to $q_c$ would be needed to observe it.

The bottom layer occupation $\rho_0$ may be related to 
$\xi ^{-1}$, where $\xi$ is the correlation length. Thus, we expect
$x_0$ to be equal to the correlation length exponent $\nu$, 
which for the KPZ equation is given by $\nu = 1/(2z - 2)$, 
yielding $x_0 = 1$ in $1d$. However, the numerical results suggest
that this scaling argument is valid only for $p>1$, whereas for 
$0<p<1$ much larger values for $x_0$ are obtained. 
This may be related to the existence of
$\em different$ universality classes in both cases, corresponding to
the distinction between an `upper' and a `lower' wall in 
Ref.~\cite{Hwa}.

In order to verify this picture we made a 
numerical estimate of the effective
nonlinear KPZ term corresponding to 
the RSOS model considered in this work.
This is done by comparing the growth velocities 
of a flat and a tilted interface in
absence of a wall. We find that the 
nonlinearity is indeed non vanishing
in the $(p,q)$ plane, except on a particular line
(the dashed line in Fig.~1).
As expected, this line and the phase transition line are different and
intersect in the point $q=p=1$. At all 
other points on the transition line
the nonlinear term is {\em not} vanishing, and the KPZ-like exponents
are expected to be valid. Since $\lambda$ is small
in the vicinity of the $q=p=1$ point, very slow crossover
phenomena occur, making it difficult to observe the true exponents
in this region.
Moreover, the interaction between the wall and the interface
strongly depends on the sign of $\lambda$. This may lead to
to different exponents $x_0$ for the bottom layer occupation
on the two sides of the special point.

{\em Acknowledgments:}  
We thank M. R. Evans for helpful discussions. 
This work was supported by Minerva Foundation, Munich, Germany.
%
%
%
%
%
%

\end{document}